\documentclass{svproc}

\usepackage[table]{xcolor}
\usepackage{times}
\usepackage{url}
\usepackage{graphicx}
\usepackage{amsmath}
\usepackage{listings}
\usepackage{color}

\definecolor{dkgreen}{rgb}{0,0.6,0}
\definecolor{gray}{rgb}{0.5,0.5,0.5}
\definecolor{mauve}{rgb}{0.58,0,0.82}

\begin{document}
\mainmatter              
\title{Gliders2d: Source Code Base for \\ RoboCup 2D Soccer Simulation League}
\titlerunning{Gliders2d: Source Code Base}  
%
\author{Mikhail Prokopenko$^{1,2}$ \and Peter Wang$^{2}$}
\authorrunning{Prokopenko and Wang} 
\institute{Complex Systems Research Group, Faculty of Engineering and IT\\ 
			The University of Sydney, NSW 2006, Australia\\
\email{mikhail.prokopenko@sydney.edu.au}\\
\and
Data Mining, CSIRO Data61, PO Box 76, Epping, NSW 1710, Australia\\
}

\maketitle              

\begin{abstract}
 We describe Gliders2d, a base code release for Gliders, a soccer simulation team which won the RoboCup Soccer 2D Simulation League in 2016. We trace six evolutionary steps, each of which is encapsulated in a sequential change of the released code, from v1.1 to v1.6, starting from agent2d-3.1.1 (set as the baseline v1.0).  These changes improve performance by adjusting the agents' stamina management, their pressing behaviour and the action-selection mechanism, as well as their positional choice in both attack and defense, and enabling riskier passes.  The resultant behaviour, which is sufficiently generic to be applicable to physical robot teams, increases the players' mobility and achieves a better control of the field. The last presented version, Gliders2d-v1.6, approaches the strength of Gliders2013, and outperforms agent2d-3.1.1 by four goals per game on average. The sequential improvements demonstrate how the methodology of human-based evolutionary computation can markedly boost the overall performance with even a small number of controlled steps. 
\end{abstract}

\section{Introduction}

The RoboCup Soccer 2D Simulation League contributes to the overall RoboCup initiative, sharing its inspirational Millennium challenge: producing a team of fully autonomous humanoid soccer players capable of winning a soccer game against the 2050 FIFA World Cup holder, while complying with the official FIFA rules \cite{Burkhard}. Over the years, the 2D Simulation League made several important advances in autonomous decision-making  under  constraints, flexible tactical planning, collective behaviour and teamwork, communication and coordination, as well as opponent modelling and adaptation \cite{Noda2003,ATAL2000,IAAI2000,Butler2001,Reis2001,Prokopenko02,Prokopenko03,Prokopenko16,Zuparic2017}.  These advances are to a large extent underpinned by the standardisation of many low-level behaviours, world model updates and debugging tools, captured by several notable base code releases, offered by ``CMUnited'' team from Carnegie Mellon University (USA)  \cite{StoneABFKLSTW00,LNAI99-simulator}, ``UvA Trilearn'' team from University of Amsterdam (The Netherlands) \cite{Kok03robocup}, ``MarliK'' team from University of Guilan (Iran) \cite{marlik}, and ``HELIOS'' team  from AIST Information Technology Research Institute (Japan) \cite{agent2d}. The latter release in 2010 included a number of components:
\begin{itemize}
\item \emph{librcsc-4.0.0}: a base library for the RoboCup Soccer Simulator (RCSS);
\item \emph{agent2d-3.0.0}: a base source code for a team;
\item \emph{soccerwindow2-5.0.0}: a viewer and a visual debugger program for RCSS;
\item \emph{fedit2-2.0.0}: a team formation editor for \emph{agent2d}.
\end{itemize}
As a result, almost 80\% of the League's teams eventually switched their code  base to agent2d over the next few years \cite{Prokopenko16}. The 2016 champion team, Gliders2016 \cite{gliders2016tdp,Prokopenko16}, was also based on the well-developed code base of \emph{agent2d-3.1.1} \cite{agent2d}, and fragments of MarliK source code \cite{marlik}, all written in C++.

The winning approach developed by Gliders combined human innovation and artificial evolution, following the methodologies of \emph{guided self-organisation} \cite{nehaniv2005evolutionary,prokopenko06-alife,sab06,prokopenko2013guided} and \emph{human-based evolutionary computation} (HBEC). The latter comprises a set of evolutionary computation techniques that incorporate human innovation \cite{kosorukoff2001human,Cheng2004}. This fusion allowed us to optimise several components, including an action-dependent evaluation function proposed in Gliders2012 \cite{POWH12}, a particle-swarm based self-localisation method and tactical interaction networks introduced in Gliders2013 \cite{gliders2013tdp,budden2013particle,liz10d,cliff2013towards,CliffLWWOP17}, a new communication scheme and dynamic tactics with Voronoi diagrams utilised by Gliders2014 \cite{gliders2014tdp}, bio-inspired flocking behaviour incorporated within Gliders2015 \cite{gliders2015tdp}, and opponent modelling diversified in Gliders2016 \cite{gliders2016tdp}. The overall framework achieved a high level of tactical proficiency ensuring players' mobility and the overall control over the soccer field. 

In this paper, we describe a base code release for Gliders, called \emph{Gliders2d}, version v1, with 6 sequential changes which correspond to 6 evolutionary HBEC steps, from v1.1 to v1.6.  Since Gliders2d release is based on agent2d, the version Gliders2d-v1.0 is identical to agent2d-3.1.1 (apart from the team name), but every next step includes a new release. It is important to point out that Gliders2d is an evolutionary branch separate from the (Gliders2012 --- Gliders2016) branch. Thus, the final version of the presented release, Gliders2d-v1.6, is neither a subset not superset of any of Gliders2012 --- Gliders2016 teams. However, as a point of reference, we note that Gliders2d-v1.6 has a strength approaching that of Gliders2013 \cite{gliders2013tdp}, and future releases will improve the performance further. 

Our objectives in making this first release are threefold: (a) it includes several important code components which explain and exemplify various approaches taken and integrated within the champion team Gliders2016; (b) it illustrates the HBEC methodology by showing some of the utilised primitives, while explicitly tracing the resultant performance (i.e., the fitness) for each sequential step from v1.1 to v1.6; (c) it demonstrates how one can make substantial advances, starting with the standard agent2d code, with only a small number of controlled steps. It may also serve as a brief tutorial that may help new teams in making the first steps within the league, using the available base code.

\vspace{-3mm}
\section{Methodology and Results}
\vspace{-1mm}

The HBEC approach evolves performance across an artificial ``generation'', using an automated evaluation of the fitness landscape, while the team developers innovate and recombine various behaviours. The mutations are partially automated.  On the one hand, the development effort translates human expertise into novel behaviours and tactics. On the other hand, the automated evaluation platform, utilised during the development of Gliders, and Gliders2d in particular, leverages the power of modern supercomputing in exploring the search-space. 

Each solution, represented as the team source code, can be interpreted as a ``genotype'', encoding the entire team behaviour in a set of ``design points''. A design point, in the context of a data-farming experiment, describes a specific combination of input parameters  \cite{Cioppa2007}, defining either a single parameter  (e.g., pressing level), complex multi-agent tactics (e.g., a set of conditional statements shaping a positioning scheme for several players), or multi-agent communication protocols \cite{Prokopenko16,Zuparic2017,gabelcommunication}. 

While some design points are easy to vary, others may be harder to mutate and/or recombine due to their internal structure.  For example, a specific tactic (design point), created by a team developer, may be implemented via several conditional statements each of which comprises a condition and an action, involving multiple parameters and primitives (see next subsections for examples). These components can then be mutated and recombined as part of the genotype. 

The solutions are evaluated against a specific opponent, over thousands of games played for each generation. In order to maintain coherence of the resultant code, which evolves against different opponents in parallel,  auxiliary conditions  switch the corresponding parts of design points on and off for specific opponents \cite{Prokopenko16}, in an analogy to epigenetic programming \cite{Tanev2008}.    The fitness function is primarily based on the average goal difference, with the average points as a tie-breaker, followed by the preference for a lower standard error.

The main thread in the evolutionary branch described in this release aims to ensure a better control of the soccer field, by different means: (i) stamina management with higher dash power rates; (ii) more intense pressing of the ball possessing opponent; (iii) actions' evaluation aimed at delivering the ball to points stretching the opposition most; (iv) attacking players positioning to maximise their ball reachability potential; (v) defending players positioning to minimise the ball reachability potential of the opponents; (vi) risky passes. These improvements may in general be applied to robotic teams in physical RoboCup leagues.

All the changes in Gliders2d are marked with 
\begin{lstlisting}
// G2d: <brief comment>
\end{lstlisting}
For example, setting the role of the agent based on its uniform number is done as follows:
\begin{lstlisting}
// G2d: role
		int role = Strategy::i().roleNumber( wm.self().unum() );
\end{lstlisting}
while retrieving the opponent name is achieved in this fashion:
\begin{lstlisting}
// G2d: to retrieve opp team name
		bool helios2018 = false;
		if (wm.opponentTeamName().find("HELIOS2018") != std::string::npos)
				helios2018 = true;
\end{lstlisting}

In tracing the relative performance of Gliders2d from v1.1 to v1.6 we used three benchmark teams: agent2d-3.1.1 itself \cite{agent2d}, Gliders2013 \cite{gliders2013tdp}, and the current world champion team, HELIOS2018 \cite{helios18}. For each sequential step, 1000 games were played against the benchmarks.  Against agent2d, the goal difference achieved by Gliders2d-v1.6 improves from zero to $4.2$.  Against HELIOS2018, the goal difference improves from $-12.73$ to $-4.34$. Finally, against Gliders2013, the goal difference improves from $-5.483$ to $-0.212$, achieving near-parity. Tables \ref{tel1}, \ref{tel2}, and \ref{tel3} summarise the performance dynamics, including the overall points for and against, goals scored and conceded, the goal difference, and the standard error of the mean.

\subsection{Gliders2d v1.1: Stamina management}

The first step in improving upon agent2d performance, along the released evolutionary branch, is adding adjustments to the agents' stamina management (confined to a single source file \verb|strategy.cpp|). Specifically, there are four additional assignments of the maximal dash power in certain situations:
\begin{lstlisting}
// G2d: run to offside line
		else if	( wm.ball().pos().x > 0.0
							&& wm.self().pos().x < wm.offsideLineX()
							&& fabs(wm.ball().pos().x - wm.self().pos().x) < 25.0
						)
							dash_power = ServerParam::i().maxDashPower();

// G2d: defenders
		else if	( wm.ball().pos().x < 10.0
							&& (role == 4 || role == 5 || role == 2 || role == 3)
						)
							dash_power = ServerParam::i().maxDashPower();

// G2d: midfielders
		else if	( wm.ball().pos().x < -10.0
							&& (role == 6 || role == 7 || role == 8)
						)
							dash_power = ServerParam::i().maxDashPower();

// G2d: run in opp penalty area
		else if	( wm.ball().pos().x > 36.0
							&& wm.self().pos().x > 36.0
							&& mate_min < opp_min - 4
						)
							dash_power = ServerParam::i().maxDashPower();
\end{lstlisting}
				
This fragment of the source code demonstrates how these specific situations are described through conditions constraining the ball position, the agent position and its role, the offside line, and the minimal intercept cycles for the Gliders2d team (\verb|mate_min|) and the opponent team (\verb|opp_min|). 

Such constraints can be evolved by mutation or recombination of primitives  (\verb|argument (op) X|), where \verb|X| is a constraint,  \verb|wm.ball().pos().x| is the argument, and \verb|(op)| is a relational operator, e.g., $<$, $>$, $==$, and so on. The action form may vary from a simple single assignment (the maximal dash power in this case), to a block of code.

Adding these four conditions increased the goal difference against HELIOS2018 from -12.729 to -6.868, and against Gliders2013 from -5.483 to -2.684.

\vspace{-2mm}
\subsection{Gliders2d v1.2: Pressing}

The second step along this evolutionary branch is adding adjustments to the agents' pressing behaviour (confined to a single source file \verb|bhv_basic_move.cpp|). 
The \verb|pressing| level is expressed as the number of cycles which separate the minimal intercept cycles by the agent (\verb|self_min|) and the fastest opponent (\verb|opp_min|). More precisely, the intercept behaviour forcing the agent to press the opponent with the ball is triggered when \verb|self_min < opp_min + pressing|. In agent2d the pressing level is not distinguished as a variable, being hard-coded as 3 cycles, and making it an evolvable variable is an example of a simple innovation. Specifically, there are four assignments of the pressing level, tailored to different opponent teams, agent roles and their positions on the field, as well as the ball location:
\begin{lstlisting}
// G2d: pressing
		int pressing = 13;

		if ( role >= 6 && role <= 8 && wm.ball().pos().x > -30.0 
					&& wm.self().pos().x < 10.0 )
					pressing = 7;

		if ( fabs(wm.ball().pos().y) > 22.0 && wm.ball().pos().x < 0.0 
					&& wm.ball().pos().x > -36.5 && (role == 4 || role == 5) ) 
					pressing = 23;

		if (helios2018)
					pressing = 4;

		if ( ! wm.existKickableTeammate()
					&& ( self_min <= 3
								|| ( self_min <= mate_min
											&& self_min < opp_min + pressing )
							)
				)
		{
				Body_Intercept().execute( agent );
				...
		}
\end{lstlisting}

Again, adding these four evolved conditions increased the goal difference against agent2d from near-zero to 1.288, against HELIOS2018 from -6.868 to -6.476 (this increase is within the standard error of the mean), and against Gliders2013 from -2.684 to -1.147.

\subsection{Gliders2d v1.3: Evaluator}

The third step modifies the action evaluator, following the approach introduced in Gliders2012 \cite{POWH12}, which diversified the single evaluation metric of agent2d by considering multiple points as desirable states. The action-dependent evaluation mechanism is described in detail in \cite{POWH12,gliders2016tdp}, and the presented release includes its implementation (source files \verb|sample_field_evaluator.cpp| and \verb|action_chain_graph|).

In particular, a new variable, \verb|opp_forward|, is introduced, counting the number of non-goalie opponents in a sector centred on the agent and extending to the points near the opponent's goal posts:  
\begin{lstlisting}
// G2d: number of direct opponents
		int opp_forward = 0;

		Vector2D egl (52.5, -8.0);
		Vector2D egr (52.5, 8.0);
		Vector2D left = egl - wm.self().pos();
		Vector2D right = egr - wm.self().pos();

		Sector2D sector(wm.self().pos(), 0.0, 10000.0, left.th(), right.th());

		for ( PlayerPtrCont::const_iterator of = wm.opponentsFromSelf().begin();
					of != wm.opponentsFromSelf().end(); ++of )
		{
					if ( sector.contains( (*of)->pos() ) && !((*of)->goalie()) )
							opp_forward++;
		} 
\end{lstlisting}
The single evaluation metric of agent2d is invoked when there are no opponents in this sector, or when the ball is located within (or close to) the own half:
\begin{lstlisting}
		if ( wm.ball().pos().x < depth || opp_forward == 0 )
		{
			// stay with best point = opp goal
		}
\end{lstlisting}
Otherwise, the logic enters into a sequence of conditions (marked in the released code), identifying the ``best'' point out of several possible candidates offered by Voronoi diagrams. A Voronoi diagram is defined as the partitioning of a plane with $n$ points into $n$ convex polygons, so that each polygon contains exactly one point, while every point in the given polygon is closer to its central point than any other \cite{DFL08}. The best point is selected to be relatively close to the teammates' positions, and far from the opponents' positions.  The distance between the identified best point and the future ball location, attainable by the action under consideration, is chosen as the evaluation result:
\begin{lstlisting}
		double weight = 1.0;
		if (wm.ball().pos().x > 35.0)
				weight = 0.3;

		double point = state.ball().pos().x * weight;
		...
		point += std::max( 0.0, 40.0 - best_point.dist( state.ball().pos() ) );
\end{lstlisting}
The condition scaling the initial assignment of the point's value, by \verb|weight|, is another example of a simple mutation. 

The action-dependent evaluation mechanism increased the goal difference against agent2d from 1.288 to 1.616, while not providing a notable improvements against the two other benchmarks, as it is applicable in attacking situations which are rare in these match-ups at this stage.

\subsection{Gliders2d v1.4: Positioning}

To make a better use of the new field evaluator, the positioning scheme of the players is adjusted by selecting points according to suitably constructed Voronoi diagrams. For example, a Voronoi diagram may partition the field according to the positions of the opponent players; the candidate location points can be chosen among  Voronoi vertices, as well as among the points located at intersections between Voronoi segments and specific lines, e.g., offside line; subject to certain constraints, as illustrated in \cite{gliders2014tdp}.  All the constrained conditions are evolvable. A small fragment of the new code, fully contained in source file \verb|strategy.cpp|, is below:
\begin{lstlisting}
// G2d: Voronoi diagram
		...
		VoronoiDiagram vd;
		...
		std::vector<Vector2D> OffsideSegm_tmpcont;

		for ( PlayerPtrCont::const_iterator o = wm.opponentsFromSelf().begin();
					o != wm.opponentsFromSelf().end();  ++o )
		{
				...
				vd.addPoint((*o)->pos());
		}
		...
		vd.compute();
		...
		for ( VoronoiDiagram::Segment2DCont::const_iterator 
					p = vd.segments().begin(), end = vd.segments().end(); 
					p != end; ++p )
		{
				Vector2D si = (*p).intersection( offsideLine );
				if (si.isValid() && fabs(si.y) < 34.0 && fabs(si.x) < 52.5)
				{
						OffsideSegm_tmpcont.push_back(si);
				}
		}
\end{lstlisting}
Once the container with the candidate points is filled, some of the players (three forwards) are assigned to the most promising points.

The positioning based on Voronoi diagrams increased the goal difference against agent2d from 1.616 to 2.387, again maintaining the performance against the two other benchmarks.

\subsection{Gliders2d v1.5: Formations}

This step did not change any of the source code files --- instead the formation files, specified in configurations such as \verb|defense-formation.conf|, \verb|offense-formation.conf|, etc. were modified with fedit2. This approach, pioneered in the Simulation League by \cite{Akiyama2008,Akiyama2010}, is based on Constrained Delaunay Triangulation (CDT) \cite{chew1989constrained}. For a set  of points in a plane, a Delaunay triangulation  achieves an outcome such that no point from the set is inside the circumcircle of any  triangle.  Essentially, CDT divides the soccer field into a set of triangles, based on the set of predefined ball locations, each of which is mapped to the positions of each player. Moreover, when the ball takes any position \emph{within} a triangle, each player's position is dynamically adjusted during the runtime in a congruent way \cite{Akiyama2008,Akiyama2010,Prokopenko16}.  Overall, a formation defined via CDT is an ordered list of coordinates, and so, in terms of evolutionary computation, mutating and recombining such a list can be relatively easily automated and evaluated. 

\begin{figure}[ht]
\vspace{-5mm}
\centering
\includegraphics[width=13.2cm]{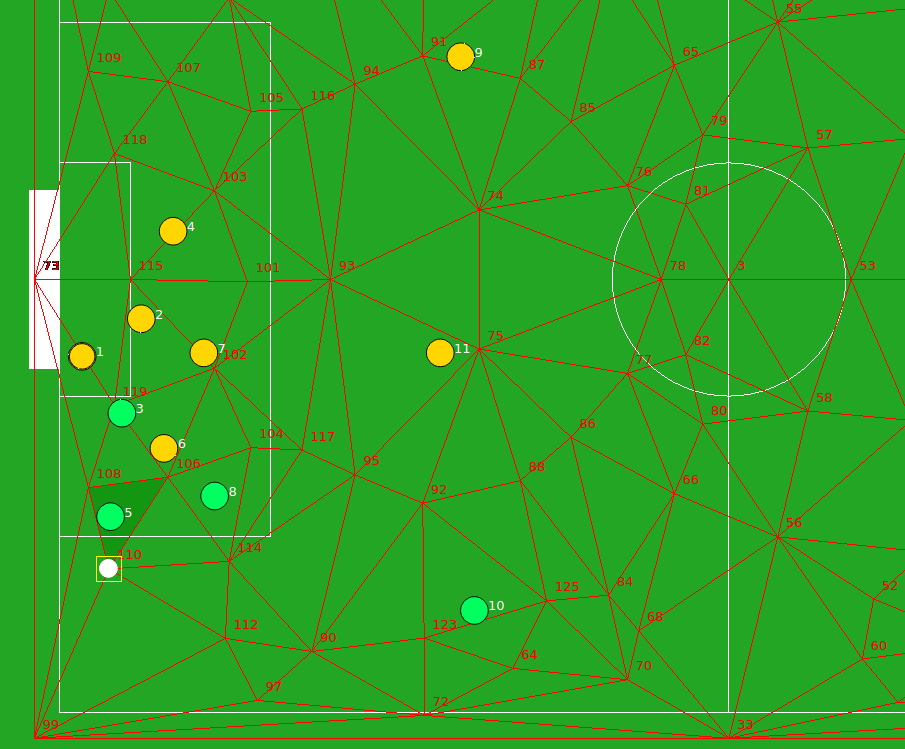}
\caption{Example of a Delaunay triangulation, used by \texttt{defense-formation.conf}, produced by fedit2. The triangle formed by points 106, 108 and 110 is highlighted. When the ball is located at 110, the players are supposed to be located in the shown positions.} 
\label{fig1}
\vspace{-5mm}
\end{figure}

Figure \ref{fig1} shows a CDT fragment; for example, the point 110, where the ball is located, defines the following intended positions for the players:
\begin{lstlisting}
Ball -48.66 22.71
1 -50.72 6.07
2 -46.08 3.12
3 -47.6 10.53
4 -43.58 -3.75
5 -48.49 18.65
6 -44.3 13.29
7 -41.17 5.8
8 -40.32 17.03
9 -21.01 -17.44
10 -19.94 26.01
11 -22.62 5.8
\end{lstlisting}

The released changes in Gliders2d-1.5 formations are aimed at improving the defensive performance, placing the defenders and midfielders closer to the own goal.  A notable performance gain was observed against all three benchmarks. The goal difference against agent2d increased from 2.387 to 3.210; against HELIOS2018: from -6.422 to -4.383; and against Gliders2013: from -1.039 to -0.344.
 
\vspace*{-3mm}
\subsection{Gliders2d v1.6: Risky passes}

The final step presented in this release introduced \verb|risk| level, expressed as the number of additional cycles ``granted'' to teammates receiving a pass, under a pressure from opponent players potentially intercepting the pass (\verb|strict_check_pass_generator.cpp|). If \verb|risk| level is set to zero, the default passing behaviour of agent2d is recovered. For positive values of \verb|risk| the passes are considered as feasible even if an ideal opponent interceptor gets to the ball trajectory sooner than the intended recipient of the pass. 
\begin{lstlisting}
// G2d: risk passes
		int risk = 0;

		if ( wm.ball().pos().x < wm.offsideLineX()
					&& receive_point.x >  wm.offsideLineX() + 3.0
					&& wm.offsideLineX() - receiver.player_->pos().x < 5.0 )
		{
				if (heliosbase)
						risk = 5;
				else if (helios2018)
						risk = 0;
				else
						risk = 2;
		}

		if ( M_pass_type == 'T' )
		{
				if ( o_step + risk <= step )
				{
						...
						failed = true;
				}
				...
		}
		else
		{
				if ( o_step + risk <= step + ( kick_count - 1 ) )
				{
						failed = true;
				}
		}
// G2d: risk in opponent check
		int risk = 0;

		if ((receive_point.x < pass_max_x || fabs(receive_point.y) > pass_min_y)
					&& (M_pass_type == 'T' || M_pass_type == 'L')
					&& fabs(ball_move_angle.degree() - oppDir) > pass_cut
					&& fabs(ball_move_angle.degree()) < pass_angle
					&& wm.ball().pos().x < wm.offsideLineX()
					&& receive_point.x >  wm.offsideLineX() + pass_depth )
		{
				if (heliosbase)
						risk =  2;
				else
						risk =  1;
		}	

		int n_step = ( n_turn == 0
									? n_turn + n_dash + risk
									: n_turn + n_dash + 1 ); // 1 step penalty for observation delay
\end{lstlisting}
The conditional statements in this fragment include several new variables, such as \verb|pass_max_x|, \verb|pass_min_y|, \verb|pass_cut|, \verb|pass_angle|, \verb|pass_depth|, used in mutating and recombining the conditions.

The addition of risky passes increased the goal difference against agent2d from 3.210 to 4.2; and against Gliders2013: from -0.344 to -0.212.

\begin{table}[ht]
\vspace*{-3mm}
\begin{center}
\begin{tabular}{|c|c|c|c|c|c|c|}
 \hline
 \scriptsize{ \ Gliders2d \ } &  \scriptsize{ \ Points for \ } &  \scriptsize{ \ Points against  \ } & \scriptsize{ \ Goals scored \ } & \scriptsize{ \ Goals conceded \ } & \scriptsize{ \ Goal diff. \ } & \scriptsize{ \ Std. error \ }  \\ \hline
 \scriptsize{ \ v0.0 (agent2d) \ } &  \scriptsize{ \ 1.384 \ } &  \scriptsize{ \ 1.414  \ } & \scriptsize{ \ 2.287 \ } & \scriptsize{ \ 2.289 \ } & \scriptsize{ \ -0.002 \ } & \scriptsize{ \ 0.040 \ }  \\ \hline
 \scriptsize{ \ v1.1 (stamina) \ } &  \scriptsize{ \ 1.345 \ } &  \scriptsize{ \ 1.468  \ } & \scriptsize{ \ 2.254 \ } & \scriptsize{ \ 2.290 \ } & \scriptsize{ \ -0.036 \ } & \scriptsize{ \ 0.049 \ }  \\ \hline
 \scriptsize{ \ v1.2 (pressing) \ } &  \scriptsize{ \ 2.161 \ } &  \scriptsize{ \ 0.691  \ } & \scriptsize{ \ 2.642 \ } & \scriptsize{ \ 1.355 \ } & \scriptsize{ \  1.288 \ } & \scriptsize{ \ 0.051 \ }  \\ \hline
 \scriptsize{ \ v1.3 (evaluator) \ } &  \scriptsize{ \ 2.252 \ } &  \scriptsize{ \ 0.607  \ } & \scriptsize{ \ 2.997 \ } & \scriptsize{ \ 1.381 \ } & \scriptsize{ \  1.616 \ } & \scriptsize{ \  0.063 \ }  \\ \hline
 \scriptsize{ \ v1.4 (positioning) \ } &  \scriptsize{ \ 2.515 \ } &  \scriptsize{ \ 0.367  \ } & \scriptsize{ \ 3.849 \ } & \scriptsize{ \ 1.461 \ } & \scriptsize{ \  2.387 \ } & \scriptsize{ \  0.086 \ }  \\ \hline
 \scriptsize{ \ v1.5 (formations) \ } &  \scriptsize{ \  2.785 \ } &  \scriptsize{ \ 0.154  \ } & \scriptsize{ \ 3.995 \ } & \scriptsize{ \ 0.785 \ } & \scriptsize{ \  3.210 \ } & \scriptsize{ \  0.181 \ }  \\ \hline
 \scriptsize{ \ v1.6 (risky passes) \ } &  \scriptsize{ \  2.840 \ } &  \scriptsize{ \  0.116  \ } & \scriptsize{ \ 5.214 \ } & \scriptsize{ \ 1.014 \ } & \scriptsize{ \  4.200 \ } & \scriptsize{ \  0.172 \ }  \\ \hline
\end{tabular}
\end{center}
\caption{Performance evaluation for Gliders2d against agent2d, over $\sim \! \! 1000$ games carried out for each version of Gliders2d against the opponent. The goal difference improves from zero to $4.2$, while the average game score improves from $(2.29 : 2.29)$ to $(5.21 : 1.01)$.}
\label{tel1}
\vspace*{-6mm}
\end{table}

\begin{table}[ht]
\vspace*{-3mm}
\begin{center}
\begin{tabular}{|c|c|c|c|c|c|c|}
 \hline
 \scriptsize{ \ Gliders2d \ } &  \scriptsize{ \ Points for \ } &  \scriptsize{ \ Points against  \ } & \scriptsize{ \ Goals scored \ } & \scriptsize{ \ Goals conceded \ } & \scriptsize{ \ Goal diff. \ } & \scriptsize{ \ Std. error \ }  \\ \hline
 \scriptsize{ \ v0.0 (agent2d) \ } &  \scriptsize{ \ 0.000 \ } &  \scriptsize{ \ 3.000  \ } & \scriptsize{ \  0.123 \ } & \scriptsize{ \ 12.852 \ } & \scriptsize{ \ -12.729 \ } & \scriptsize{ \ 0.514 \ }  \\ \hline
 \scriptsize{ \ v1.1 (stamina) \ } &  \scriptsize{ \  0.001 \ } &  \scriptsize{ \ 2.998  \ } & \scriptsize{ \ 0.231 \ } & \scriptsize{ \ 7.099 \ } & \scriptsize{ \ -6.868 \ } & \scriptsize{ \ 0.276 \ }  \\ \hline
 \scriptsize{ \ v1.2 (pressing) \ } &  \scriptsize{ \  0.003 \ } &  \scriptsize{ \ 2.994  \ } & \scriptsize{ \ 0.248 \ } & \scriptsize{ \  6.724 \ } & \scriptsize{ \  -6.476 \ } & \scriptsize{ \  0.140 \ }  \\ \hline
 \scriptsize{ \ v1.3 (evaluator) \ } &  \scriptsize{ \  0.004 \ } &  \scriptsize{ \ 2.992  \ } & \scriptsize{ \ 0.269 \ } & \scriptsize{ \ 6.821 \ } & \scriptsize{ \  -6.552  \ } & \scriptsize{ \  0.310 \ }  \\ \hline
 \scriptsize{ \ v1.4 (positioning) \ } &  \scriptsize{ \ 0.002 \ } &  \scriptsize{ \ 2.996  \ } & \scriptsize{ \ 0.298 \ } & \scriptsize{ \ 6.720 \ } & \scriptsize{ \ -6.422 \ } & \scriptsize{ \  0.223 \ }  \\ \hline
 \scriptsize{ \ v1.5 (formations) \ } &  \scriptsize{ \  0.027 \ } &  \scriptsize{ \ 2.952  \ } & \scriptsize{ \ 0.273 \ } & \scriptsize{ \  4.655 \ } & \scriptsize{ \   -4.383 \ } & \scriptsize{ \  0.197 \ }  \\ \hline
 \scriptsize{ \ v1.6 (risky passes) \ } &  \scriptsize{ \  0.024 \ } &  \scriptsize{ \   2.961  \ } & \scriptsize{ \ 0.260 \ } & \scriptsize{ \ 4.600 \ } & \scriptsize{ \  -4.337 \ } & \scriptsize{ \  0.161 \ }  \\ \hline
\end{tabular}
\end{center}
\caption{Performance evaluation for Gliders2d against HELIOS2018, over $\sim \! \! 1000$ games carried out for each version of Gliders2d against the opponent. The goal difference improves from $-12.73$ to $-4.34$, while the average game score improves from $(0.12 : 12.85)$ to $(0.26 : 4.60)$.}
\label{tel2}
\vspace*{-6mm}
\end{table}

\begin{table}[ht]
\vspace*{-3mm}
\begin{center}
\begin{tabular}{|c|c|c|c|c|c|c|}
 \hline
 \scriptsize{ \ Gliders2d \ } &  \scriptsize{ \ Points for \ } &  \scriptsize{ \ Points against  \ } & \scriptsize{ \ Goals scored \ } & \scriptsize{ \ Goals conceded \ } & \scriptsize{ \ Goal diff. \ } & \scriptsize{ \ Std. error \ }  \\ \hline
 \scriptsize{ \ v0.0 (agent2d) \ } &  \scriptsize{ \ 0.022 \ } &  \scriptsize{ \  2.968  \ } & \scriptsize{ \ 0.569 \ } & \scriptsize{ \  6.052 \ } & \scriptsize{ \ -5.483 \ } & \scriptsize{ \ 0.213 \ }  \\ \hline
 \scriptsize{ \ v1.1 (stamina) \ } &  \scriptsize{ \ 0.183 \ } &  \scriptsize{ \ 2.730  \ } & \scriptsize{ \  0.596 \ } & \scriptsize{ \ 3.280 \ } & \scriptsize{ \  -2.684 \ } & \scriptsize{ \ 0.071 \ }  \\ \hline
 \scriptsize{ \ v1.2 (pressing) \ } &  \scriptsize{ \ 0.539 \ } &  \scriptsize{ \ 2.230  \ } & \scriptsize{ \ 0.613 \ } & \scriptsize{ \ 1.760 \ } & \scriptsize{ \   -1.147 \ } & \scriptsize{ \  0.063 \ }  \\ \hline
 \scriptsize{ \ v1.3 (evaluator) \ } &  \scriptsize{ \ 0.657 \ } &  \scriptsize{ \ 2.109  \ } & \scriptsize{ \ 0.770 \ } & \scriptsize{ \ 1.800 \ } & \scriptsize{ \   -1.030 \ } & \scriptsize{ \  0.067 \ }  \\ \hline
 \scriptsize{ \ v1.4 (positioning) \ } &  \scriptsize{ \ 0.603 \ } &  \scriptsize{ \ 2.160  \ } & \scriptsize{ \ 0.700 \ } & \scriptsize{ \ 1.739 \ } & \scriptsize{ \ -1.039  \ } & \scriptsize{ \  0.077 \ }  \\ \hline
 \scriptsize{ \ v1.5 (formations) \ } &  \scriptsize{ \  1.039 \ } &  \scriptsize{ \ 1.607  \ } & \scriptsize{ \ 0.700 \ } & \scriptsize{ \ 1.044 \ } & \scriptsize{ \  -0.344 \ } & \scriptsize{ \   0.026 \ }  \\ \hline
 \scriptsize{ \ v1.6 (risky passes) \ } &  \scriptsize{ \  1.111 \ } &  \scriptsize{ \  1.527  \ } & \scriptsize{ \ 0.776 \ } & \scriptsize{ \  0.988 \ } & \scriptsize{ \   -0.212 \ } & \scriptsize{ \  0.038 \ }  \\ \hline
\end{tabular}
\end{center}
\caption{Performance evaluation for Gliders2d against Gliders2013, over $\sim \! \! 1000$ games carried out for each version of Gliders2d against the opponent. The goal difference improves from  $-5.48$ to $-0.21$, while the average game score improves from $(0.57 : 6.05)$ to $(0.78 : 0.99)$.}
\label{tel3}
\vspace*{-16mm}
\end{table}

\section{Conclusions}

In this paper, we described the first version of \emph{Gliders2d}: a base code release for Gliders (based on agent2d-3.1.1). We trace six sequential changes aligned with six evolutionary steps.  These steps improve the overall control of the pitch by increasing the players' mobility through several means: less conservative usage of the available stamina balance (v1.1); more intense pressing of opponents (v1.2); selecting more diversified actions (v1.3); positioning forwards in open areas (v1.4);  positioning defenders closer to own goal (v1.5); and considering riskier passes (v1.6). 

As has been argued in the past, the simulation leagues enable replicable and robust investigation of complex robotic systems \cite{RAM15,ProkopenkoWMBLC17}. We believe that the purpose of the RoboCup Soccer Simulation Leagues (both 2D and 3D) should be to simulate agents based on a futuristic robotic architecture which is not yet achievable in hardware. Aiming at such a general and abstract robot architecture may help to identify a standard for what humanoid robots may look like in 2050, the year of the RoboCup Millennium challenge. This is the reason for focussing, in this release, on the features which can also be used by simulated 3D, as well as robotic, teams competing in RoboCup, aiming at some of the most general questions: when to conserve energy (stamina), when to run (pressing), where to kick the ball (actions), where to be on the field (positioning in attack and defense), and when to take risks (passes).  While the provided specific answers may or may not be widely acceptable,  general reasoning along these lines may bring us closer to a new RoboCup Humanoid Simulation League (HSL). In HSL, the Simulated Humanoid should be defined in a standard and generalisable  way, approaching human soccer-playing behavior \cite{PHILOSOC10}, while the behavioural and tactical improvements can be evolved and/or adapted to this standardised architecture.

The location of the released code: \verb|http://www.prokopenko.net/gliders2d.html|. 

The last presented version, Gliders2d-v1.6, is comparable to Gliders2013, achieving the average score  of $(0.78 : 0.99)$ against this benchmark, and outperforms agent2d-3.1.1 with the average score  $(5.21 : 1.01)$.

In tracing this evolutionary branch, we illustrated the methodology of human-based evolutionary computation, showing that even a small number of controlled steps can dramatically improve the overall team performance. 

\section{Acknowledgments}
We thank several members of Gliders team contributing during 2012--2016: David Budden, Oliver Cliff, Victor Jauregui and Oliver Obst.  We are also grateful to participants of the discussion on the future of the RoboCup Simulation Leagues, in particular to Peter Stone, Patrick MacAlpine, Nuno Lau, Klaus Dorer, and Daniel Polani.

\bibliographystyle{splncs}

\end{document}